\pdfoutput=1
\documentclass[CEJP,PDF]{cej} % use PDF command to enable PDFLaTeX driver
\usepackage{indentfirst}
\usepackage{layout}
\usepackage{amsmath}
\usepackage{textcomp}
\usepackage{hyperref}
\usepackage{graphicx}

\title{Optical properties and Mechanical properties of C, Si, Ge and 3C-SiC Materials Calculated from First Principles Theory}

\articletype{Research Article}

\author{Xuejie Liu\inst{1},
        Liangfang Li\inst{1}$^,$\email{liliangfangllf@163.com},
        Feng Lu\inst{1}}

\institute{
    \inst{1} School of Mechanical Engineering, Inner Mongolia University of Science \& Technology,\\
     Baotou, Inner Mongolia 014010, PR China
          }

\abstract{First-principles calculations based on Density functional theory (DFT) and Heyd, Scuseria and Ernzerhof (HSE) adopt PBE approximation-a new version of the generalized gradient approximation (GGA). It has studied lattice parameter, electronic structure, optical properties and mechanical properties of materials, including diamond-C and zinc-blende structure of Si, Ge, 3C-SiC. The result of HSE calculation is obviously superior to DFT calculation and accord well with the existing experimental values. It indicates that they are indirect bandgap materials from the figure of band structure and density of states. The bandgap which it calculated using HSE accords well with the existing experimental values except Ge. In the calculation of optical property, it shows that they correspond with the existing experimental values from the imaginary part of analytical dielectric function to the refractive index and adsorption coefficient. It demonstrates that they exists the corresponding relationship among the peak of the imaginary part of analytical dielectric function, the refractive index and adsorption coefficient. The optical property has a direct relationship with the distribution of crystal bandgap and electronic state density. From the results of mechanical properties it can know that they are brittle materials. Though the hardness and stiffness of 3C-SiC is lower than diamond, it is superior to the materials of Si, Ge as excellent semiconductor materials. In addition, the mechanical anisotropy of four materials is inconspicuous; the anisotropy of Diamond-C is very inconspicuous from anisotropic properties of Young modulus.}


\begin{document}
\maketitle

%% ###################################################################

\section{Introduction}

Diamond has high hardness and bulk modulus [1,2]. It has been widely applied in the mold abrasive and related machinery industry. In addition, with its excellent optical performance, it has become an important part of optical element in the synchrotron [3,4]. Si is main semiconductor materials and has taken a leading position in semiconductor materials. As a kind of important semiconductor materials, Ge has high carrier mobility. It is widely used in the field of infrared optical materials, the transistor, the solar cell, radiation detector and thermoelectric materials and so on [5]. With its character of wide band-gap, high breakdown critical field strength, high electron mobility and strong ability to resist radiation and so on, SiC has a broad prospect of application [6,7].
The mechanical properties and optical properties of diamond, semiconductor and its basic optical function play an important role in the application of materials and devices of semiconductor. Firstly, DFT was used to optimize crystal structure in this paper; Secondly, it calculated and compared the bandgap diversity of DFT [8] with HSE [9-11]; Then it used HSE to calculate basic optical function, such as dielectric function and reflectivity, studied the relations between them; Finally, it calculated the mechanics parameters and made a simple analysis.

\section{Calculation method and details}

In this work, C is diamond structure; Si, Ge and SiC are zinc-blende structures. It used VASP programs [12] within PAW method [13] to conduct the optical properties and mechanical properties calculations. It used 7x7x7 Monkhorst k-point mesh. The exchange-correlation functional is used GGA [14-16] of PBE, PBEsol, B3 formalism [12,17-19] in DFT and HSE calculations. The lattice parameters  of above materials are shown in table 1. In optical properties and mechanical properties calculations, it uses PBEsol formalism in HSE calculations. The same Monkhorst k-point mesh was used in HSE calculations, but it added to 2-3 times k-point mesh and bands in optical properties calculations. The electron wave function is expanded in plane waves up to a cutoff energy of 500 eV. All of ionic relaxations were performed with a tolerance of 0.000001 eV.

In optical properties calculations, the frequency dependent imaginary part of the dielectric function ${\varepsilon_2}\left(\omega\right)$ is determined by a summation over empty states using the following equation [20]:

\begin{align}
&\varepsilon _{\alpha \beta }^{\left( 2 \right)}\left( \omega  \right) = \frac{{4{\prod ^2}{e^2}}}{\Omega }\frac{1}{{{q^2}}}\mathop {\lim }\limits_{q \to 0} \varepsilon _{\alpha \beta }^{\left( 2 \right)}\left( \omega  \right) = \frac{{4{\prod ^2}{e^2}}}{\Omega }\frac{1}{{{q^2}}}\mathop {\lim }\limits_{q \to 0}  \times {\left\langle {{u_{ck + {e_\alpha }q}}\left| {{u_{vk}}} \right.} \right\rangle ^*}
\end{align}

Where the indices c and v refer to conduction and valence band states respectively, and ${u_{ck}}$ is the cell periodic part of the wavefunctions at the k-point k. The real and imaginary parts of analytical dielectric function are connected by the Kramers-Kronig relation as:

\begin{align}
&\varepsilon _{\alpha \beta }^{\left( 1 \right)}\left( \omega  \right) = 1 + \frac{2}{\prod }P\int_0^\infty  {\frac{{\varepsilon _{\alpha \beta }^{\left( 2 \right)}\left( {\omega '} \right)\omega '}}{{\omega {'^2} - {\omega ^2} + i\eta }}} d\omega '
\end{align}

Where P denotes the principle value.
The Refractive index and Adsorption coefficient formulae are presented below.

\begin{align}
&n = {\left[ {\left( {\sqrt {\varepsilon _1^2 + \varepsilon _2^2}  + {\varepsilon _1}} \right)/2} \right]^{\frac{1}{2}}}
\end{align}

\begin{align}
&\alpha  = {\left[ {2\left( {\sqrt {\varepsilon _1^2 + \varepsilon _2^2}  - {\varepsilon _1}} \right)} \right]^{\frac{1}{2}}}
\end{align}

In mechanical properties calculations, the Shear modulus, Bulk modulus, Young's modulus and Poisson's ratio are given below [21].

\begin{align}
&G = \frac{1}{2}\left( {{c_{11}} - {c_{12}}} \right)
\end{align}

\begin{align}
&B = \frac{1}{3}\left( {{c_{11}} + 2{c_{12}}} \right)
\end{align}

\begin{align}
&E = \frac{{9BG}}{{3B + G}}
\end{align}

\begin{align}
&\mu  = \frac{{3B - 2G}}{{6B + 2G}}
\end{align}

\section{Calculation results and discussion}

\subsection{Electronic structure}

It know that the accuracy of the lattice parameter and bandgap directly affect the optical properties calculations. So, in this work, the structural optimizations were performed within PBE, PBEsol and B3 approximation. It used experimental lattice constant which are presented in table 1. It chose PBEsol formalism which is very close to experimental lattice to the follow calculations.

% Table generated by Excel2LaTeX from sheet 'Sheet3'
\begin{table}[htbp]
  \centering
  \caption{PBE, PBEsol and B3 lattice were compared with experimental lattice constant.}
    \begin{tabular}{lllll}
    \hline
    {Symbol} &{Lattice PBE({\AA})} & {Lattice PBEsol({\AA})} & {Lattice B3({\AA})} & {Lattice Exp({\AA})[22]} \\
    \hline\hline
    Diamond-C & 3.575 & 3.564 & 3.552 & 3.567 \\
    Si    & 5.47  & 5.436 & 5.443 & 5.43 \\
    Ge    & 5.784 & 5.737 & 5.801 & 5.658 \\
    3C-SiC & 4.38  & 4.363 & 4.349 & 4.36 \\
    \hline
    \end{tabular}%
  \label{tab:addlabel}%
\end{table}%

In this section it used HSE06 approximations to calculate the DOS (Figure 1) and band structures (Figure 2). By means of the PBE and HSE06 approximation it compared bandgap of Diamond-C, Si, Ge and 3C-SiC which was given in table 2. In HSE06 approximation, it found that bandgaps are excellent for wide range of semiconductors, except for very large gap systems.

% Table generated by Excel2LaTeX from sheet 'Sheet2'
% Table generated by Excel2LaTeX from sheet 'Sheet5'
\begin{table}[htbp]
  \centering
  \caption{PBE band gap and HSE06 band gap were compared with experimental values.}
    \begin{tabular}{llllll}
    \hline
    Symbol & Structure & Band-gap Type & PBE Band-gap & HSE Band-gap & Reference[22] \\
  \hline\hline
    Diamond-C & Diamond & Indirect & 4.12  & 5.34  & 5.4 \\
    Si    & Zincblende & Indirect & 0.58  & 1.23  & 1.17 \\
    Ge    & Zincblende & Indirect & 0.01  & 0.39  & 0.74 \\
    3C-SiC & Zincblende & Indirect & 1.37  & 2.35  & 2.36 \\
   \hline
    \end{tabular}%
  \label{tab:addlabel}%
\end{table}%

Figure 1(a) shows the TDOS and PDOS of Diamond-C. The low-energy area of Valence band is mainly constituted by C 2s and C 2p orbital electrons. The high-energy area is mainly constituted by C 2p and C 2d orbital electrons. Figure 1(b) gives the TDOS and PDOS of Si. The low-energy area of Valence band is mainly constituted by Si 3s, Si 3p and Si 3d orbital electrons. The High-energy area is mainly constituted by Si 3p and Si 3d orbital electrons. Figure 1(c) shows the TDOS and PDOS of Ge. The low-energy area of Valence band is mainly formed by Ge 4s, Ge 4p and Ge 4d orbital electrons. The low-energy area of valence band is mainly formed by Ge 4p and Ge 4d orbital electrons. From top to bottom in figure 1(d), it listed the PDOS of Si and C of 3C-SiC and the TDOS of 3C-SiC. The Valence band is composed of two sub bands. The sub band of low-energy area is mainly composed of Si 3s, Si 3p and C 2s orbital electrons. The sub band of high-energy area is mainly made up of Si 3s, Si 3p, Si 3d and C 2s, C 2p orbital electrons. In the sub band of high-energy area, the low-energy area is mainly composed of Si 3s and C 2p orbital electrons and the high-energy area is mainly composed of Si 3p and C 2p orbital electrons.

\begin{figure}[h]
\centering
\includegraphics[width=16cm]{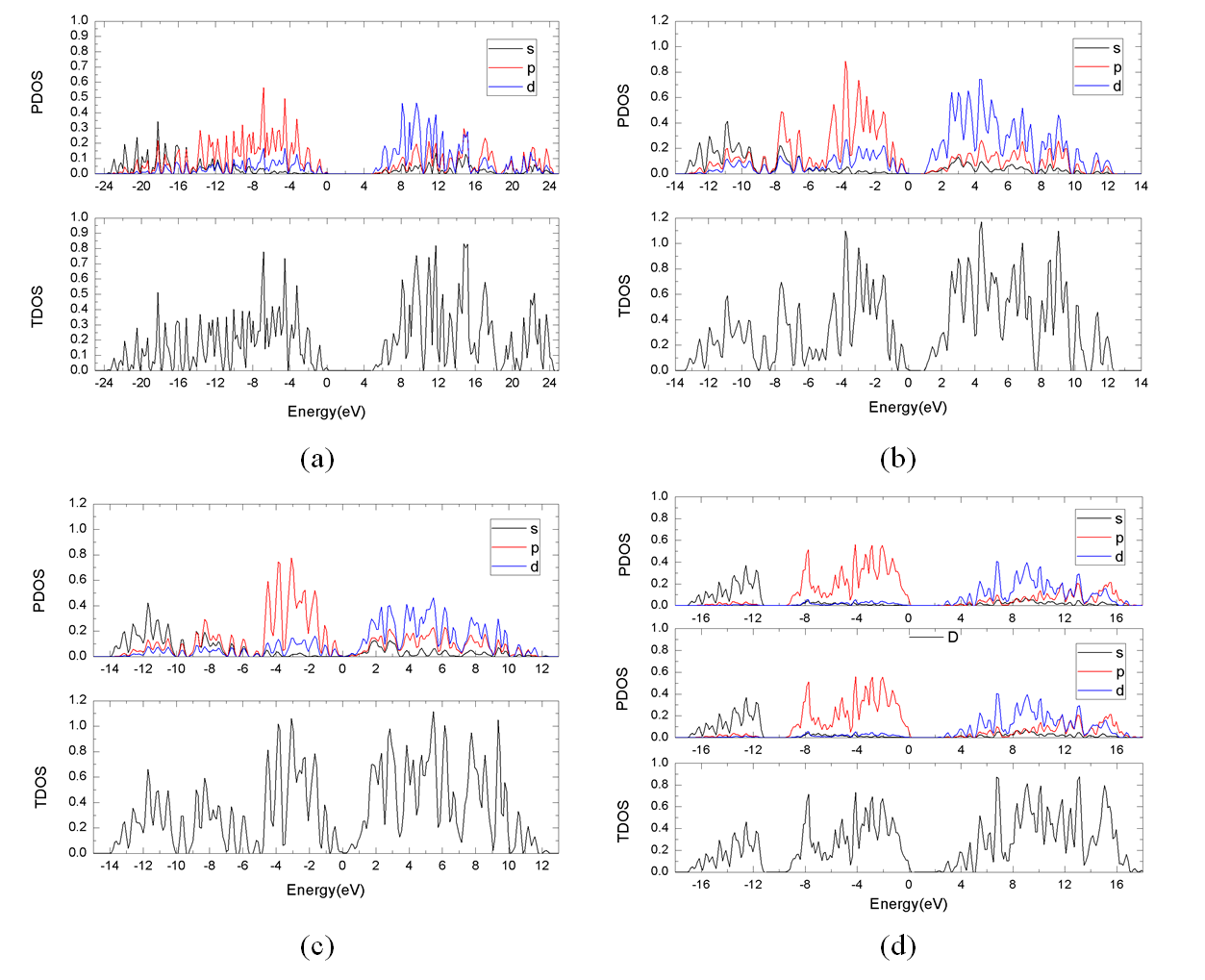}
\caption{Total density of states (TDOS) and angular momentum projected partial density of states (PDOS) of (a) Diamond-C, (b) Si, (c) Ge and (d) 3C-SiC.}
\end{figure}

Figure 2(a) gives the band structure of Diamond-C, it indicates that the top position of valence band lie on the ¦£ point of Brillouin zone, the bottom position of conduction band locate between the X and the ¦£ point of Brillouin zone. The indirect bandgap, Eg=5.34eV, accords with the experimental value, Eg =5.40eV. Figure 2(b) shows the band structure of Si. it shows that the top position of valence band lie on the ¦£ point of Brillouin zone, the bottom position of conduction band locate between the X and the ¦£ point of Brillouin zone. The indirect bandgap, Eg=1.23eV, is close to the experimental value, Eg=1.17eV. It can find that the top position of valence band lie on the ¦£ point of Brillouin zone, the bottom position of conduction band locate between the X and the ¦£ point of Brillouin zone from the Figure 2(c). The indirect bandgap, Eg=0.39eV, has large difference comparing to the experimental value, Eg =0.74eV. There is the bond structure of 3C-SiC; it indicates the typical characteristics of bond of the zinc-blende structure with indirect band gap. it shows that the top position of valence band lie on the ¦£ point of Brillouin zone, the bottom position of conduction band lie on the X point of Brillouin zone. The indirect bandgap, Eg=2.35eV, accords with the experimental value, Eg =2.36eV.

\begin{figure}[h]
\centering
\includegraphics[width=16cm]{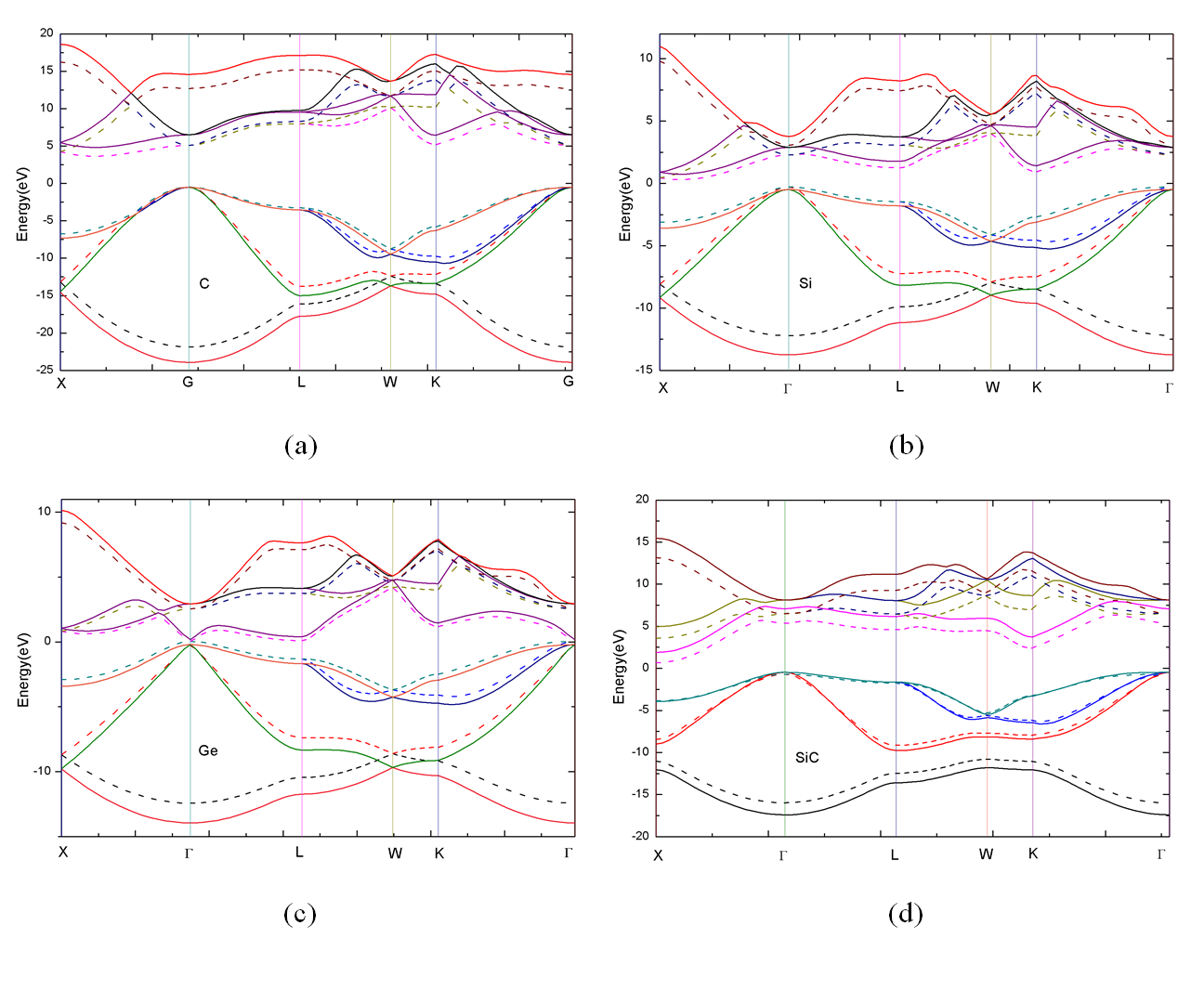}
\caption{The band structure of Si, Ge, 3C-SiC in zinc blende structures and Diamond -C in Diamond structures are shown here. The solids lines and dashed lines represent PBE and HSE calculations respectively. Calculated HSE band gaps of Diamond -C, Si, 3C-SiC are close to experimental value. Calculated PBE and HSE band-gap of Ge are incomparable with experimental, but Calculated HSE band-gap of Ge is much better.}
\end{figure}

\subsection{Optical properties}

As a bridge to connect with transition microscopic physical process of the band and solid electronic structure, the dielectric function reflects the solid energy band structure and several of spectral information. Diamond-C, Si, Ge are spacing band gap, but the 3C-SiC is direct bandgap semiconductor. The optical spectrum of 3C-SiC is formed by electron transition between energy level. Each dielectric peak can be explained through the band structure and density of states.

It could directly get dielectric function matrix by conclusions and extracted the data to make dielectric function diagram, as shown in figure 3. It respectively listed the real part and imaginary part of Diamond-C, Si, Ge, 3C-SiC of dielectric function in figure 3(a) and 3(b). According to the formula 3 and 4, it can derive respectively refractive index and adsorption coefficient rate, as shown in figure 3(c) and 3(d). From the figure 3(b) it can see the Diamond-C, Si, Ge and 3C-SiC respectively have two peaks, the corresponding photon energy respectively are 8.29eV, 12.98eV; 3.70eV, 4.45eV; 1.74eV, 4.43eV; 8.22eV, 10.002eV.

 \begin{figure}[h]
\centering
\includegraphics[width=16cm]{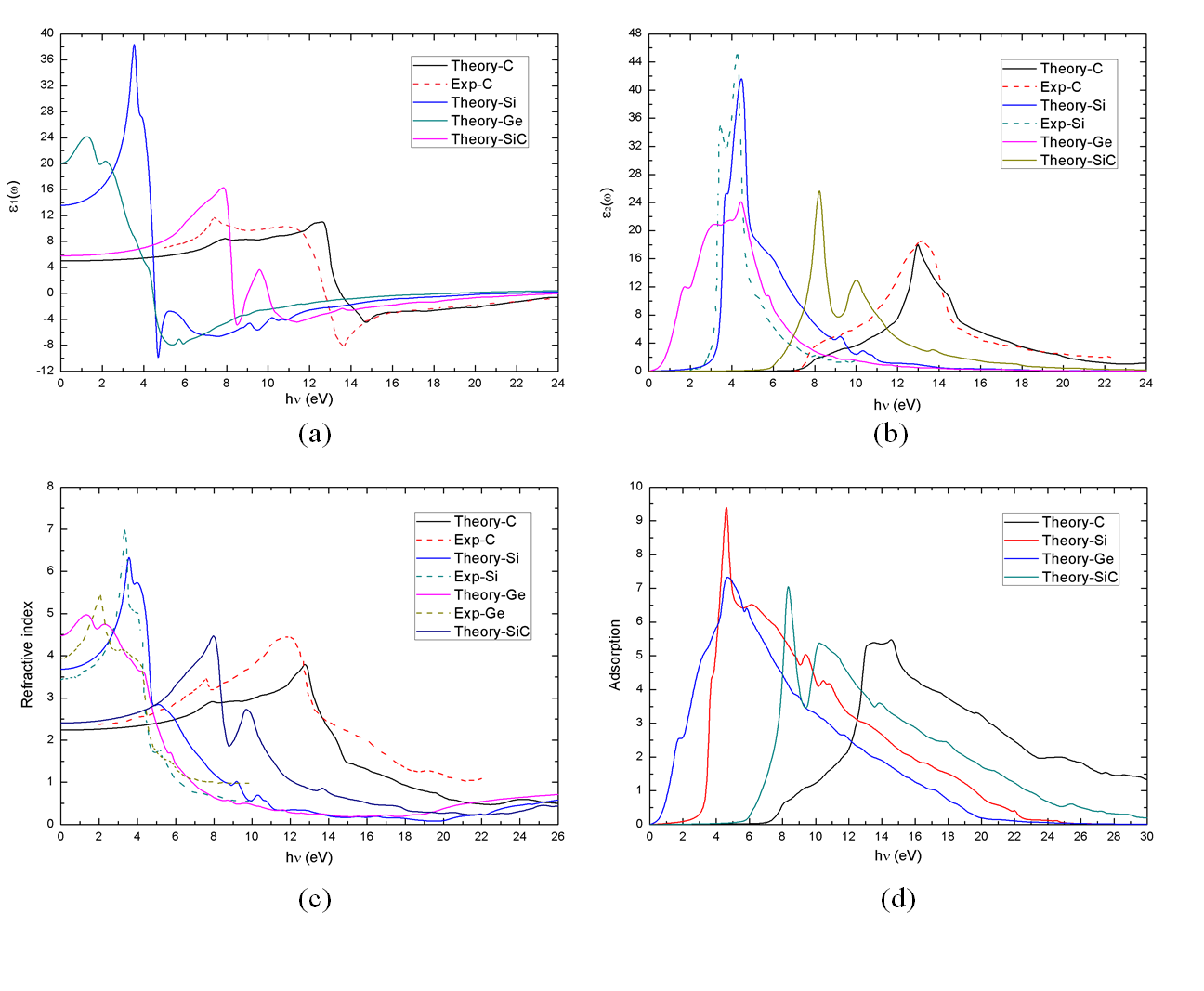}
\caption{Calculated real and imaginary part of dielectric functions, Refractive index and Adsorption coefficient for C diamond structures and Si, Ge, 3C-SiC in zinc blende structures. Some calculation value of optical properties can be compared with experimental value.}
\end{figure}

According to density of states to analysis, Diamond-C has a weak peak in the place of 8.29eV combined with figure 1(a), there has a refraction peak in the place of 7.92eV corresponding the figure 3(c) and has an absorption peak in the place of 8.23eV corresponding the figure 3(d). The refraction peak is closely approximate to experimental measuring 7.61eV [19], the transition of electron may be transition between the electrons of C 2s and 2p or 2d orbits. Diamond-C has a dielectric peak in the position of 12.98eV and is slight deviation comparing to the experimental value 13.20eV [22], there has a refraction peak in the place of 12.74eV corresponding the figure 3(c) and has an absorption peak in the place of 13.47eV corresponding the figure 3(d). The refraction peak is slight deviation comparing to the experimental value 11.76eV [19]£¬the transition electrons mainly comes from electrons of C 2p and 2d orbits.

Corresponding the figure 1(b), Si has a small dielectric peak in the place of 3.70eV and is close to the experimental value 3.43eV [22], there has a refraction peak in the place of 3.57eV corresponding the figure 3(c) and has an absorption peak in the place of 4.61eV corresponding the figure 3(d). The refraction peak is closely approximate to experimental measuring 3.34eV [22], the transition of electronic may be transition between the electrons of Si 3p and 3d orbits. However, there appears a larger peak in the place of 4.45eV, and is slight deviation comparing to the experimental value in the position of 4.24eV [22]. Si has a refraction peak in the position of 3.99eV corresponding the figure 4(c) and has an absorption peak in the place of 6.13eV corresponding the figure 4(d). The refraction peak is in accordance with the experimental value 3.89eV [22]£¬the transition electrons mainly comes from electrons of Si 3s and 3p orbits.

As it shown in figure 3(b) and figure 2(c), Ge appears a smaller dielectric peak in the place of 1.74eV, there has a refraction peak in the place of 1.34eV corresponding the figure 3(c) and has an absorption peak in the place of 1.79eV corresponding the figure 3(d). The refraction peak has a large deviation comparing to the value of experimental measuring 2.08eV [22], Transiting electrons probably comes from electrons of Ge 4s and 4p orbits. In the place of 4.43eV, it appears a larger peak, Ge has a refraction peak in the position of 4.38eV corresponding the figure of 3(c) and has an absorption peak in the place of 4.69eV corresponding the figure 3(d). The refraction peak is in accordance with the experimental value 4.46eV [22]£¬the transition electrons may come from electrons of Ge 4p and 4d orbits.

3C-SiC has a larger peak in the place of 8.22eV in the figure 3(d) and figure 1(d), there has a refraction peak in the place of 7.79eV corresponding the figure 3(c) and has an absorption peak in the place of 8.34eV corresponding the figure 3(d), the transition electrons may come from electrons of C 2p and Si 3d orbits. But in the peak of 10.02eV, 3C-SiC has a refraction peak in the position of 9.18eV corresponding the figure of 3(c) and has a absorption peak in the place of 10.23eV corresponding the figure 3(d), the transition electrons may come from electrons of C 2s and Si 3p orbits or C 2p and C 2d orbits.

In conclusion, there has corresponding relation among the imaginary part of dielectric functions, refractive index and adsorption coefficient. It indicates that optical properties have a direct relationship with crystal band, the distribution of electronic state density.

\subsection{Mechanical properties}

It got directly Shear modulus (G), Bulk modulus (B) and C44 through the calculation. It derived Young's modulus (E), Poisson¡¯s ratio ($\mu$) , C11, C12 and C44 by formula (5-8), as it shown in table 3.

% Table generated by Excel2LaTeX from sheet 'Sheet9'
\begin{table}[htbp]
  \centering
  \caption{The Elastic constants of Diamond -C, Si, Ge and 3C-SiC.}
    \begin{tabular}{llllllll}
  \hline
    Symbol & G/GPa & B/GPa & E/GPa & $\mu$  & C11/Gpa & C12/Gpa & C44/Gpa \\
  \hline\hline
    Diamond -C & 488.783 & 406.776 & 1046.993 & 0.071 & 1058.487 & 80.921 & 581.714 \\
    Si    & 54.942 & 90.817 & 137.165 & 0.248 & 164.073 & 54.189 & 84.263 \\
    Ge    & 39.688 & 63.953 & 98.656 & 0.243 & 116.87 & 37.494 & 64.392 \\
    3C-SiC & 142.374 & 207.026 & 347.469 & 0.22  & 396.858 & 112.11 & 259.345 \\
    \hline
    \end{tabular}%
  \label{tab:addlabel}%
\end{table}%

According to the Born-Huang of lattice dynamics theory, the condition of mechanical stability was set as follows [23,24]. Diamond-C, Si, Ge and 3C-SiC all belong to cubic crystal system, the conditions of mechanics stability of cubic crystal system is,

\begin{align}
&{C_{44}} > 0,{C_{11}} > \left| {{C_{12}}} \right|,{C_{11}} + 2{C_{12}} > 0
\end{align}

From table 3 it know that the elastic constant Cij of the structure of Diamond-C, Si, Ge and 3C-SiC meet the criterion of mechanical stability and has mechanical stability.

Bulk modulus (BH) and Shear modulus (GH) both use the Voigt-Reuss-Hill approximation method [25-27], modulus calculation of cubic crystal system uses this formula (10-14):

\begin{align}
&{B_V} = {B_R} = \frac{1}{3}\left( {{C_{11}} + 2{C_{12}}} \right)
\end{align}

\begin{align}
&{G_V} = \frac{1}{5}\left( {{C_{11}} - {C_{12}} + 3{C_{44}}} \right)
\end{align}

\begin{align}
&{G_R} = \frac{{5\left( {{C_{11}} - {C_{12}}} \right){C_{44}}}}{{4{C_{44}} + 3\left( {{C_{11}} - {C_{12}}} \right)}}
\end{align}

\begin{align}
&{B_H} = \left( {{B_V} + {B_R}} \right)/2
\end{align}

\begin{align}
&{G_H} = \left( {{G_V} + {G_R}} \right)/2
\end{align}

Where Cij is elastic stiffness constant, Sij is elastic compliance tensor (mechanics matrix Sij=1/Cij). The Bulk modulus (BH) and Shear modulus (GH) which are gotten though the data calculated from formula (10-14) and table 3 are listed in table 4. From the table 4, it know that Bulk modulus (BH) and Shear modulus (GH) of Diamond-C are very high; it indicates that it¡¯s super hard. The Bulk modulus (BH) and Shear modulus (GH) of 3C-SiC also are higher, which illustrates it has higher hardness.

% Table generated by Excel2LaTeX from sheet 'Sheet10'
\begin{table}[htbp]
  \centering
  \caption{ The Bulk modulus (BV, BR, BH), Shear modulus (GV, GR, GH) and mechanical anisotropic parameters (AU) of Diamond-C, Si, Ge and 3C-SiC.}
    \begin{tabular}{lllllllll}
    \hline
    Symbol & BV/GPa & BR/GPa & GV/GPa & GR/Gpa & BH/Gpa & GH/Gpa & BH/GH & AU \\
    \hline\hline
    Diamond -C & 406.776 & 406.776 & 544.542 & 540.601 & 406.776 & 542.571 & 0.749 & 0.036 \\
    Si    & 90.817 & 90.817 & 72.535 & 69.439 & 90.817 & 70.987 & 1.279 & 0.223 \\
    Ge    & 63.952 & 63.952 & 54.51 & 51.556 & 63.952 & 53.032 & 1.206 & 0.287 \\
    3C-SiC & 207.026 & 207.026 & 212.556 & 195.197 & 207.026 & 203.876 & 1.015 & 0.445 \\
    \hline
    \end{tabular}%
  \label{tab:addlabel}%
\end{table}%

Because  BH/GH can characterize the toughness and brittleness of material, the higher BH/GH is, the better the toughness of material is. Generally speaking, BH/GH $<$ 1.75 represents material is brittle, BH/GH $>$ 1.75 represents material is ductility [29]. Two kinds of structure of the four kinds of material are brittle materials, the diamond BH/GH=0.75 is identical with BH/GH=0.8 [29], it indicates that it is very crisp fragile.

Young's modulus (E) and Poisson¡¯s ratio ($\mu$) are two important physical quantities to describe the elastic behavior of the materials. Young's modulus (E) can characterize the stiffness of materials, the stiffness of materials becomes higher with the increase of the elastic modulus; Poisson¡¯s ratio ($\mu$) can represents the ratio which the maximum tensile strength of solid material divided by the maximum shear strength, according to the criterion of fracture behavior, material of low Poisson¡¯s ratio belongs to the brittleness material. With the calculation results shown Young's modulus (E) of Diamond-C is higher than Ge, that is to say, it has a high stiffness; However, Poisson's ratio is very low ($\nu = 0.071$) , it expresses that Diamond-C is very brittle material, Poisson¡¯s ratio ($\mu$) can show the extent of the directional covalent bond. Diamond - C has very low Poisson¡¯s ratio ($\mu$) , it displays C has the very high hardness due to strong covalent bond; this is consistent with the fact. Due to the low Young's modulus (E) of Si and Ge, they have a very small stiffness. The Young's modulus (E) of 3C-SiC is between Diamond-C and Si, its stiffness also is between Diamond-C and Si. In order to characterize anisotropy of mechanical properties of materials, in this work it calculated the anisotropy parameters AU [29,30]. AU is defined as follows:

\begin{align}
&{G_H} = \left( {{G_V} + {G_R}} \right)/2
\end{align}

According to the value of AU in table 4, Diamond-C, Si, Ge and 3C-SiC all belong to anisotropic material, but the value of Diamond-C is very low, so the mechanical properties of anisotropic is most inconspicuous.

It drew anisotropic diagrams of the Young's modulus in three dimensional spaces to understand intuitively the mechanical anisotropy of Cubic crystal system Diamond-C, Si, Ge and 3C-SiC. It can find intuitively the difference of mechanical anisotropy of Diamond-C, Si, Ge and 3C-SiC. In view of the anisotropy of body elastic modulus is not generally more obvious than young's modulus, therefore it don¡¯t discuss here. It can use a similar method to draw the three dimensional anisotropic diagram for shear modulus. In this work, it uses formula (16) to draw Young's modulus of anisotropy of three dimensions.

\begin{align}
&{A^U} = 5\frac{{{G_V}}}{{{G_R}}} + \frac{{{B_V}}}{{{B_R}}} - 6 \ge 0
\end{align}

In the formula (16), l1, l2 and l3 express the cosine of angle between vector and the orthogonal coordinate axis, but in the polar coordinate, they can be expressed as ${l_1} = \sin \theta \cos \varphi$, ${l_2} = \sin \theta \sin \varphi$ and ${l_3} = \cos \theta$, it can draw the three dimensional anisotropic diagram of Young's modulus in the polar coordinate with the Mathematica5.0. The calculation formula (17) of cubic crystal which is simplified in the polar coordinate is,

\begin{align}
&\begin{array}{c}
 \frac{1}{E} = l_1^4{S_{{\rm{11}}}}{\rm{2}}l_1^2l_2^2{S_{{\rm{12}}}} + {\rm{2}}\iota l_1^2l_3^2{S_{{\rm{13}}}} + l_2^4{S_{{\rm{22}}}} + {\rm{2}}l_2^2l_3^2{S_{{\rm{23}}}}{\rm{ + }}l_3^4{S_{33}}{\rm{ + }} \\
 l_2^2l_3^2{S_{44}}{\rm{ + }}l_1^2l_3^2{S_{55}}{\rm{ + }}l_1^2l_2^2{S_{66}} \\
 \end{array}
\end{align}

The elastic compliance tensors S11, S12 and S44 which are got from the formula 18-20 are listed in table 5.

\begin{align}
&\begin{array}{c}
 \frac{1}{E} = {S_{11}}\left[ {{{\sin }^4}\theta \left( {{{\cos }^4}\varphi  + {{\sin }^4}\varphi } \right) + {{\cos }^4}\theta } \right] + \left( {2{S_{12}} + {S_{44}}} \right). \\
 \left( {{{\sin }^4}\theta {{\cos }^2}\varphi {{\sin }^2}\varphi  + {{\sin }^2}\theta {{\cos }^2}\theta } \right) \\
 \end{array}
\end{align}

\begin{align}
&{S_{11}} = \frac{{{C_{11}} + {C_{12}}}}{{\left( {{C_{11}} - {C_{12}}} \right)\left( {{C_{11}} + 2{C_{12}}} \right)}}
\end{align}

\begin{align}
&{S_{12}} = \frac{{ - {C_{12}}}}{{\left( {{C_{11}} - {C_{12}}} \right)\left( {{C_{11}} + 2{C_{12}}} \right)}}
\end{align}

\begin{align}
&{S_{44}} = \frac{1}{{{C_{44}}}}
\end{align}

% Table generated by Excel2LaTeX from sheet 'Sheet11'
\begin{table}[htbp]
  \centering
  \caption{Elastic compliance tensor.}
    \begin{tabular}{lllll}
    \hline
    Symbol & S11   & S12   & S44   & 2S12+S44 \\
    \hline\hline
    Diamond -C & 0.000955116 & -0.000067832 & 0.001719058 & 0.001583393 \\
    Si    & 0.007290466 & -0.001810040 & 0.011867605 & 0.008247525 \\
    Ge    & 0.010136231 & -0.002462035 & 0.015529879 & 0.010605809 \\
    3C-SiC & 0.002877953 & -0.000633924 & 0.003855868 & 0.002588019 \\
    \hline
    \end{tabular}%
  \label{tab:addlabel}%
\end{table}%

According to the formula (17) and the datum in the table 5, it draw out the diagram of three dimensional anisotropy of Young's modulus, from figure 4 it can find that the anisotropy of Young's modulus is very not obvious because of the sake of symmetry in the cubic crystal system, but the anisotropy of Young's modulus of Diamond-C is most inconspicuous, which is consistent with foregoing paragraphs. The anisotropy of Young's modulus of Si, Ge and 3C-SiC is obvious well.

\begin{figure}[h]
\centering
\includegraphics[width=14cm]{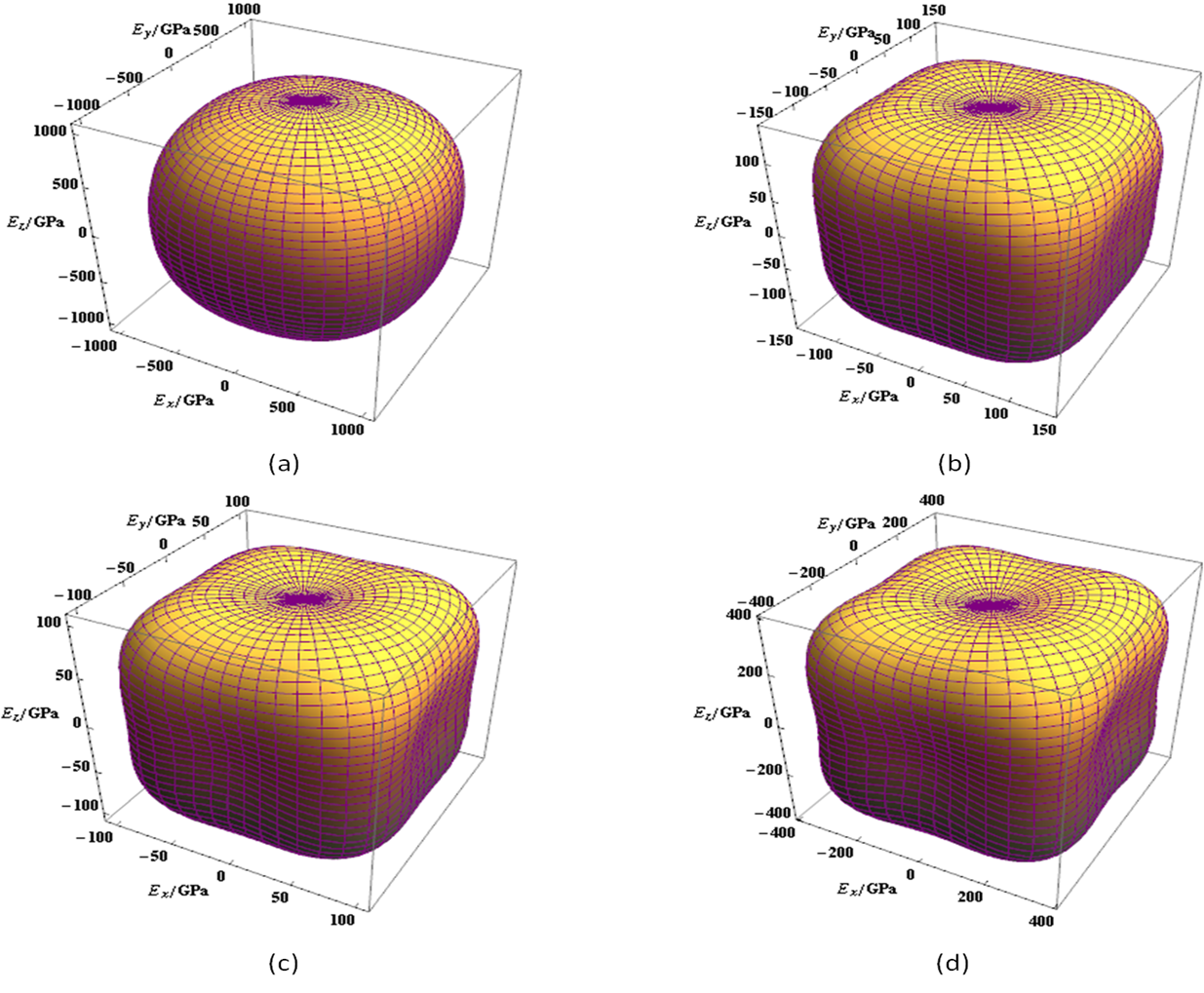}
\caption{Anisotropic properties of Young modulus for (a) Diamond-C, (b) Si, (c) Ge and (d) 3C-SiC.}
\end{figure}

\section{Conclusions}

It have calculated the electronic structure, optical properties and mechanical properties of Diamond-C, Si, Ge and 3C-SiC using the primary principle of HSE method. The bandgap calculated of Diamond-C, Si and 3C-SiC conforms to the experimental value, but Ge is less than the experimental value. It shows that the HSE method is better than the DFT method by comparison. It has calculated the imaginary part and real component of the dielectric function. It have studied structure of the critical point in the dielectric function and analysed the corresponding optical transition. According to the dielectric function it has got the relationship among the refractive index, the absorption rate and the photon energy. The calculated results of theory are basic consistent with the experiment through comparison. Through calculating and analysing the basic mechanical parameters, it finds that the stiffness of Diamond-C is the highest and the anisotropy of Young's modulus is not obvious. This is consistent with the fact. The stiffness of 3C-SiC is superior to Si and Ge, it can be used as semiconductor materials which the rigidity requirement is very high.

\section*{Acknowledgments}

For this research, it is very gratitude to the support of the National Natural Science Foundation of China (NNSFC) under grant 50845065, by Inner Mongolia Natural Science Foundation under grant 2010Zd21, by Inner Mongolia Education and Research Foundation under grant NJ06077 through the Education Ministry of Inner Mongolia, PR China.

%% ###################################################################


\begin{thebibliography}{99}

\bibitem{journal-1} Occeli F. Loubeyre P and Letoullec R. Nature materials. 2003, 2, 151
\bibitem{journal-2} Dubrovinskaia N. Dub S and Dubrovinsky L. Nano Letters. 2006, 6, 824.
\bibitem{journal-3} Guoxin M. Peng X and Shengbao Q. Vacuum Electronics. 2009, 5, 1002.
\bibitem{journal-4} Bin L. Yanli W. Journal of Henan University of Urban Construction. 2011, 4, 20.
\bibitem{journal-5} Jinhui Z. Dongli Y. Journal of Yanshan Univwersity. 2012,3, 224.
\bibitem{journal-6} Pengshou X. Chanagkun X. and Haibin P. Journal of University of Science and Technology of China. 2006, 09, 0253.
\bibitem{journal-7} Yue P. Yintang Y. Science Press. 2000, 1.
\bibitem{journal-8} P. Hohenberg and W. Kohn, Phys. Rev. B 1964, 864, 136.
\bibitem{journal-9} Ming-Zhu, Huang and W. Y. Ching, Phys. Rev. B 1993, 47, 9449.
\bibitem{journal-10} M Marsman, J Paier, A Stroppa and G Kresse. J. Phys.: Condens. Matter 20 (2008) 064201.
\bibitem{journal-11} Yoon-Suk Kim, Kerstin Hummer and Georg Kresse, Phys. Rev. B 2009, 80, 035203.
\bibitem{journal-12} http://cms.mpi.univie.ac.at/VASP/
\bibitem{journal-13} P. E. Blochl, Phys. Rev. B 1994, 50, 17953.
\bibitem{journal-14} D. C. Langreth and M. J. Mehl, Phys. Rev. B 1983, 28, 1809.
\bibitem{journal-15} J. P. Perdew and Y. Wang, Phys. Rev. B 1986, 33, 8800.
\bibitem{journal-16} D. C. Langreth and J. P. Perdew, Phys. Rev. B 1980, 21, 5469.
\bibitem{journal-17} P. Perdew, M. Ernzerhof and K. Burke, J. Chem. Phys. 1996, 105, 9982.
\bibitem{journal-18} J. P. Perdew, K. Burke and M. Ernzerhof, Phys. Rev. Lett. 1996, 77, 3865.
\bibitem{journal-19} Raghuveer Chimata. Optical Properties of Materials Calculated from First Principles Theory. 2010
\bibitem{journal-20} M. Gajdos, K. Hummer and G. Kresse, Phys. Rev B 2006, 73, 045112.
\bibitem{journal-20} C. Kittel, Introduction To Solid State Physics, 8th Edition
\bibitem{journal-22} http://www.ioffe.ru/SVA/NSM/Semicond/
\bibitem{journal-23} Patil, S. K. R. Khare, S. V. Tuttle, B. R. Bording, J. K and Kodambaka, S. Phys. Rev. B 2006, 73, 104118.
\bibitem{journal-24} Wu, Z. J. Zhao, E. J. Xiang, H. P. Hao, X. F. Liu, X. J. and Meng, J. Phys. Rev. B 2007, 76, 054115.
\bibitem{journal-25} Reuss, A. Z. Angew. Math. Mech. 1929, 9, 49.
\bibitem{journal-26} Hill, R. Proc. Phys. Soc. 1952, 65, 349.
\bibitem{journal-27} Zhao, J. J. Winey, J. M. and Gupta, Y. M. Phys. Rev. B 2007, 75, 094105.
\bibitem{journal-28} Li, Y. F. Gao, Y. M. Xiao, B. Min, T. Fan, Z. J. Ma, S. Q. and Xu, L. L. J. Alloy. Compd. 2010, 502, 28.
\bibitem{journal-29} Ravindran, P. Fast, L. Korzhavyi, P. A. Johansson, B. Wills, J. and Eriksson, O. J. Appl. Phys. 1998, 84, 4891.
\bibitem{journal-30} Ranganathan, S. I. Starzewski, M. O. Phys. Rev. Lett. 2008, 101, 055504.

\end{thebibliography}
\end{document}